\documentstyle[12pt]{article} 
\topmargin
-1.5cm 
\begin{document} 
\scrollmode

\vglue 1.7cm

\noindent {\large 
{\bf The Master Thesis of Mosh\'e Flato}}

\vspace{2.5cm}
\noindent {\bf Maurice Kibler}

\noindent {Institut de Physique Nucl\'eaire de Lyon,}

\noindent {IN2P3-CNRS et Universit\'e Claude Bernard,}

\noindent {43 Bd du 11 Novembre 1918,} 

\noindent {F-69622 Villeurbanne Cedex, France}

   \vspace{2cm}

\noindent {to my Master and Friend Mosh\'e Flato}

   \setcounter{page}{0}

\vspace{2cm}
\noindent {\bf Abstract}

\noindent 
The genesis and impact of the M.Sc.~thesis by Mosh\'e Flato is
analysed. In this connection, the fruitful passage of Mosh\'e in Lyon, {\it
capitale mondiale de la gastronomie}, is evoked. Finally, some basic elements 
for a model in crystal-field theory are given as a 
important step on the way opened by Mosh\'e.

\vspace{2.4 cm}

\noindent {Key words~: finite subgroups of $SU(2)$, Wigner-Racah algebra for 
finite  and  compact groups, crystal-field theory, molecular and condensed 
matter spectroscopy.}

\vspace{0.6 cm}


\vspace{1.8 cm}

\noindent {Contributed   paper  submitted  for  the  proceedings 
of the `Conf\'erence Mosh\'e Flato 1999' held in Dijon (France), 
5--8 September 1999. The proceedings will be published by Kluwer 
Academic Publishers in the series `Mathematical Physics Studies',
eds.~G.~Dito and D.~Sternheimer, 
founded by the late Professor Mosh\'e Flato.}


   \newpage

       \baselineskip 0.685  true cm 

\noindent {\bf Geneses}
\vspace{0.3cm}

\noindent Mosh\'e Flato prepared his M.Sc.~thesis under the 
guidance of Giulio Racah at the Physics Department of the Hebrew University 
in Jerusalem at the end of the fifties \cite{F60}.
The final examination took place in 1960. The genesis of the thesis
can be described as follows. 

From 1942 to 1949, Racah published a series of four major papers 
on new algebraic and group theoretical methods for the 
analysis of complex spectra in atomic and nuclear 
spectroscopy [2-5]. In particular, he developed the
concepts of 
  {\it irreducible tensor operator}, 
  {\it coefficient of fractional parentage} 
  (a concept which goes back to Goudsmit and Bacher) and 
  {\it seniority number}. Furthermore, he 
showed how the use of a
chain of continuous groups ending with $SO(3) \supset SO(2)$
makes it possible to calculate the coefficients of fractional parentage and to
classify the state vectors originating from an $\ell^N$ configuration
in spherical symmetry. In 1956, he became very 
interested in crystal-field theory (a theory for partly-filled shell
ions embedded in crystals, solutions or complexes) on the occasion 
of discussions and seminars with Wily Low from 
the Microwave Division of the Department of 
Physics of the Hebrew University \cite{BCde-SST}. Crystal-field 
theory, also referred to as  ligand-field theory,  
may be considered as a branch of quasi-atomic physics in the
sense that this theory is concerned with the electronic properties
of a central ion (with a configuration of type $\ell^N$ where 
$\ell = d$ or $f$) in interaction with its immediate surrounding.      
Racah realized that the simultaneous consideration of finite groups
(for describing the ion in its environment) and continuous groups
(for labelling the Russell-Saunders terms $[S,L]$ or multiplets
$[J]$ of the ion) can be very fruitful in crystal-field theory.  
W.~Low and G.~Racah suggested to two students, Miss G.~Schoenfeld
and Mr.~G.~Rosengarten, to calculate the Coulomb, spin-orbit and
crystal-field matrices for $d^N$ ions in cubical symmetries. The
cubical symmetry is often an idealized (or first-order) symmetry so
that G.~Racah asked Mr.~M.~Flato to reconsider the problem of 
$d^N$ ions in finite symmetry and to treat the case of trigonal and
tetragonal symmetries which are of interest in absorption
spectroscopy  and  in electron spin resonance spectroscopy.

Schoenfeld calculated the matrices for the $d^2$ and $d^3$ 
configurations in cubical
symmetry (symmetry group $G = O$) and Rosengarten obtained the
matrices  for the $d^4$ and $d^5$ 
configurations in cubical
symmetry too. Flato dealt with the case where distortions
(i.e., {\it deformations}) of trigonal symmetry (symmetry group 
$G = D_3$) or tetragonal symmetry (symmetry group $G = D_4$) are 
introduced  in order to describe a greater number of 
physical situations. Along these lines, he calculated the
crystal-field matrices  for the $d^2$ and $d^3$ 
configurations in tetragonal and trigonal symmetries. 

{}From the group-theoretical point of view, the problem addressed by Schoenfeld and
Rosengarten, and in a more complete way by Flato, concerns the labelling of the
state vectors arising from the $d^N$ configuration via the use of a chain of
groups starting from the classification group $U(5)$ and ending with the point
symmetry group $G$ or its spinor group $G^*$ with $G \sim G^*/Z_2$. The head group
$U(5)$ corresponds to the fact that the $2 d + 1 = 5$ orbitals
associated to the $d$--shell (for which $\ell=2$) may undergo a unitary
transformation without changing the spectrum of the relevant Hamiltonian. The
end group is $G$ (the geometrical symmetry group for the ion in its 
environment) or $G^*$ (the spectral group
that labels the spin, called the double group of $G$ in the terminology of
H.A.~Bethe) 
depending on the parity, even or odd,  of the number of 
electrons $N$. Therefore, Flato used the chains 
$$
U(5) \supset SO(3) \supset O \supset D_3 \ {\rm or} \ D_4 
$$
and
$$
U(5) \supset SU(2) \supset O^* \supset D_3^* \ {\rm or} \ D_4^* 
$$           
for the configurations $d^2$ and $d^3$, respectively. 

One of the  advantages  in
using such chains of groups is that the Coulomb and spin-orbit matrices already
calculated for the free ion can be easily 
transferred to the case of the ion in its
surrounding. Indeed, the chain $U(5) \supset SO(3)$ or $U(5) \supset SU(2)$ is
useful  for  the construction of the Coulomb and spin-orbit
matrices. From a qualitative point of view, the chains 
$SO(3) \supset O   \supset D_3   \ {\rm or} \ D_4$ and 
$SU(2) \supset O^* \supset D_3^* \ {\rm or} \ D_4^*$ serve to describe the 
(external) 
{\it symmetry breaking} (arising from an inhomogeneous Stark effect)
when going from the free ion to the ion in its
environment. From a quantitative point of view, the calculation of the
crystal-field matrix can be done via the Wigner-Eckart theorem for the group
$SU(2)$ in a nonstandard basis, namely a basis adapted to the chain 
$SO(3) \supset O   \supset D_3   \ {\rm or} \ D_4$    for the $d^2$ configuration 
and the chain 
$SU(2) \supset O^* \supset D_3^* \ {\rm or} \ D_4^*$  for the $d^3$
configuration. 

To fully understand the originality of the works by 
Flato,  one has ro realize  that at this time most of the works on crystal- and
ligand-field theories for $d^N$ ions in finite symmetries (involving both an 
idealized or high symmetry and a distortion or low symmetry) were conducted
according to the approach by 
Tanabe, Sugano and Kamimura 
(from the Japanese school of
M.~Kotani) [7-10]
and the one by Griffith [11-18].  
In these approaches, the  
end group, i.e., the point symmetry group
$G$ or its spinor group $G^*$, is considered as an isolated group 
so that the subduction (or descent in symmetry)  
$$
SO(3) \supset H   \supset G \ {\rm or} \ SU(2) \supset H^* \supset G^*
$$
is not fully taken into account ($H$ stands for a high symmetry group, like 
$O$, and $G$ for a low symmetry group, like $D_3$)~: the information about the
descent in symmetry $H   \supset G$  or  $H^* \supset G^*$ is kept 
at the minimal level and 
the link with $SO(3)$ or $SU(2)$ is not  completely  exploited. 
Therefore, in the Griffith  [11-18] and 
Tanabe, Sugano and Kamimura [7-10]  approaches, all the necessary matrix elements must
be calculated through the use of the Wigner-Racah algebra of $G$ or $G^*$.  
As a consequence, the calculation of the crystal-field energy matrix is very
easy but the calculation of the Coulomb and spin-orbit energy matrices is very
complicated. As a matter of fact, the invariance of the Coulomb and spin-orbit
interactions under $SO(3)$, when translated in terms of $H \supset G$, 
leads to  sums  of invariant irreducible tensor operators 
which are difficult to handle via the Wigner-Eckart theorem
for the group $G$ or $G^*$. This difficulty does not appear in the approach 
followed by
Schoenfeld, Rosengarten and Flato. 

The M.Sc.~work by Flato \cite{F60} constituted the first systematic attempt to combine
the simplifications  afforded  by the various   chains of groups  of the type
$U(5) \supset SO(3) \supset H   \supset G  $  and  
$U(5) \supset SU(2) \supset H^* \supset G^*$ for dealing with 
partly-filled shell ions in discrete symmetry. The works by Schoenfeld,
Rosengarten and Flato were used by Low and several of his collaborators 
 for understanding 
optical spectra and electron paramagnetic 
resonance spectra of transition-metal ions in crystals 
(for example, see Refs.~[6,19-21]). Although 
Mosh\'e presented his M.Sc.~thesis dissertation in 1960, several
years later the 
subject was still up to date since he published a part of his thesis in 
1965 \cite{F65}.   

The 1965 Flato paper and some impulse given by Mosh\'e 
opened an avenue of new investigations in~: (i) the
Wigner-Racah algebra of the group $SU(2)$ in bases adapted to chains of the
type $SU(2) \supset G^*$ where $G^*$ is a group or a chain of groups of interest
in molecular physics and condensed matter physics and (ii) the development of
models in optical spectroscopy (mainly luminescence spectroscopy and
photoemission spectroscopy) of $d^N$ and $f^N$ ions in crystalline fields 
(see Refs.~[23-32] and references therein). 
Along this vein, in 1966 Mosh\'e Flato suggested to a student, Maurice Kibler,
from the `Facult\'e des Sciences de l'Universit\'e de Lyon', to extend his
M.Sc.~thesis work to the situation where a magnetic interaction 
(arising from the Zeeman effect) is added to the
Coulomb + spin-orbit + crystal-field Hamiltonian. This suggestion led the 
present author
to develop the Wigner-Racah algebra of an arbitrary finite group embedded in
$SU(2)$ and to construct a phenomenological model, where symmetry
considerations play a central role, for ions in a crystal subjected to an
external magnetic field \cite{K68,K69IJQC,K69CRAS}. The research in that direction continued during several
years with some students and collaborators of Kibler, especially in~: 

- the study of specific chains of groups, 

- the application of symmetry adaptation 
  to electron paramagnetic resonance, 

- the unification of crystal-field and
  ligand-field theories,  
  
- the development of models for 
  photo-electron spectroscopy and 
  
- the derivation of a phenomenological approach to
  multi-photon electronic spectroscopy. 
  
\noindent {In 1995, with the advent of strong
technological progress in two-photon absorption spectroscopy of rare earth
ions doped materials, part of the  above-mentioned works
were revived in view of the elaboration of 
a model for analysing intensities of two-photon 
transitions between states of opposite parities 
for an $\ell^N$ ion in finite symmetry \cite{DK95}.}

The rest of this paper is devoted to 
some personal recollections concerning  the  passage  of Mosh\'e in Lyon
and to
a brief description (in the appendix) of  the basic ingredients for symmetry
adaptation techniques in molecular and condensed matter spectroscopy 
developed by the present author more than thirty years ago under the direction 
of Mosh\'e.

\vspace{0.6cm} 
\noindent {\bf Mosh\'e in Lyon}
\vspace{0.3cm}
  
\noindent The arrival of Mosh\'e in Lyon in 1964 has been a great event for many people. 
Although Mosh\'e  did not yet have a   `Doctorat d'\'Etat' (he submitted 
his `th\`ese de doctorat \`es sciences physiques' at `la Sorbonne' in Paris on
June 15, 1965), he joined the university staff of the `Universit\'e de  
Lyon' as an associated professor in Physics under the recommendation of Jean
Braconnier, `doyen de la Facult\'e des Sciences de Lyon'. (Jean Braconnier 
had been a member of the Bourbaki group.) 
He kept his position of associated professor for three years, 
sharing his life between Lyon and Paris. From 1964 to 1967, 
it was a real pleasure to  
enjoy the presence in Lyon of  Mosh\'e and Daniel (Daniel Sternheimer),
 almost every Monday and Tuesday to the best of my recollection. During the academic
year 1964-65, Mosh\'e gave master lectures on group theory both for finite
groups and continuous groups. For most people in the audience the way 
to treat the subject was quite new (discussion at two levels~: 
mathematical exposure  and  illustration from  many branches of physics) and 
the lectures were very successful. The next two academic years, Mosh\'e
undertook to give a complete panorama of modern physics starting from
classical mechanics    up to quantum field theory (with intermediate steps in
relativistic mechanics, quantum mechanics and statistical mechanics). The
corresponding lectures were given by Mosh\'e with an extraordinary economy
principle in the presentation, a high level content and a constant
preoccupation of the understanding by the audience.  

During this short period (1964-67), several students (D.~Allouche, 
S.~Ekong and J.~Sternheimer) achieved their `th\`ese
de doctorat de sp\'ecialit\'e (troisi\`eme cycle)'  
and two other students (J.~Gr\'ea and M.~Kibler) started to work on
their `th\`ese d'\'Etat' under the supervision of Mosh\'e.  

Many of us in Lyon (students, collaborators and friends) have greatly benefited from 
Mosh\'e not only for a transmission of a high level knowledge but also for 
having made possible to share with him  
 a part of his dynamism, his enthusiasm, his love of physics (following 
 Christian Fr\o nsdal, one can add  he  was a physicist and a mathematician, 
 teaching mathematics and having a passion for physics) and his respect of others.   
We also learned from Mosh\'e how  to   be   demanding   for ourselves.   His 
generosity (in all aspects of life) 
and his so particular way to consider 
each of us as a unique component of the world are qualities 
which should inspire us all.
  
\vspace{0.6cm} 
\noindent {\bf Appendix~: Finite Symmetry Adaptation}
\vspace{0.3cm}

\noindent Let $G^*$ be a (finite) subgroup of $SU(2)$. We use $\Gamma$ 
($= \Gamma_0, \Gamma_1, \Gamma_2, \cdots$) to denote an irreducible
representation class (IRC) of $G^*$, $\Gamma_0$ being the identity IRC of $G^*$.
We write $[\Gamma]$ for the dimension of a matrix representation associated to
$\Gamma$. Such a matrix representation can be spanned by vectors of type
$$
\vert j a \Gamma \gamma) \; := \; \sum_{m=-j}^j \; 
\vert j m ) \; (j m \vert j a \Gamma \gamma) 
\eqno (1)
$$
which are linear combinations of vectors $\vert j m )$  adapted 
to the chain  $SU(2) \supset U(1)$. In Eq.~(1), 
the branching label $a$  has  to be used when the IRC $\Gamma$ of 
$G^*$ occurs several times in the IRC $(j)$ of $SU(2)$~; this 
external multiplicity label may be described, at least 
partially, by IRC's of subgroups of $SU(2)$ that contain in 
turn the group $G^*$. Furthermore, the label $\gamma$ is necessary when 
$[\Gamma] > 1$~; it  may be described 
sometimes by IRC's of subgroups of $G^*$.   Equation (1) is of a 
group theoretical nature~: the reduction coefficients 
$(jm \vert j a \Gamma \gamma)$ depend on the chain $SU(2) \supset G^*$ 
but not on Physics. They may be chosen, thanks to Schur's lemma, 
in such a way that the matrix representation associated to $\Gamma$ is the same
for all $j$'s and $a$'s. Equation (1) 
defines the $\left\{ j a \Gamma \gamma \right\}$ 
scheme for the chain $SU(2) \supset G^*$ that is more 
appropriate in molecular  or  condensed matter spectroscopy than the 
$\left\{ j m \right\}$ scheme for the chain $SU(2) \supset U(1)$.  
 For $j$, $a$ and $\Gamma$ fixed, the set 
$\left\{ \vert j a \Gamma \gamma) \ : \ \gamma = 1, \cdots, [\Gamma] \right\}$ 
is a  $G^*$--irreducible tensorial set  of vectors associated to $\Gamma$. 
Similarly, from the $SU(2) \supset U(1)$ spherical tensor 
operators $U^{(k)}_q$, we define the operators 
$$
U^{(k)}_{a \Gamma \gamma} \; := \; \sum_{q = -k}^k 
U^{(k)}_q \; (kq \vert k a \Gamma \gamma) 
\eqno (2)
$$
so that, for $k$, $a$ and $\Gamma$ fixed, the set 
$\left\{ U^{(k)}_{a \Gamma \gamma} \ : \ \gamma = 1, \cdots, [\Gamma] \right\}$ 
is a $G^*$--irreducible tensorial set of operators associated to $\Gamma$.
The latter $G^*$--irreducible tensorial sets are also labelled by 
IRC's of $SU(2)$ and, therefore, we can easily generate, by direct 
sum, nonstandard $SU(2)$--irreducible tensorial sets. Thus, we may 
apply the Wigner-Eckart theorem for the group $SU(2)$ in a 
nonstandard basis adapted to its subgroup $G^*$. As a result, 
we have 
         $$
    (\tau_1 j_1 a_1 \Gamma_1 \gamma_1 \vert U^{(k)}_{a \Gamma \gamma} \vert 
    \tau_2 j_2 a_2 \Gamma_2 \gamma_2) \; = \;
    (\tau_1 j_1 \Vert U^{(k)} \Vert 
    \tau_2 j_2) 
    $$
    $$
    \times
    \sum_{a_3 \Gamma_3 \gamma_3}
    \pmatrix{
                          & j_1 &                        \cr\cr
    a_3 \Gamma_3 \gamma_3 &     & a_1 \Gamma_1 \gamma_1} 
    \; 
    \overline f
    \pmatrix{
    j_1                   & k               & j_2 \cr\cr
    a_3 \Gamma_3 \gamma_3 & a \Gamma \gamma & a_2 \Gamma_2 \gamma_2}^*
    \eqno (3)
         $$
with 
         $$
   \overline f 
   \pmatrix{
   j_1                   & j_2                   & j_3\cr\cr
   a_1 \Gamma_1 \gamma_1 & a_2 \Gamma_2 \gamma_2 & a_3 \Gamma_3 \gamma_3} 
   \;  :=  \;  \sum_{m_1 m_2 m_3} \ 
   \pmatrix{
   j_1 & j_2 & j_3
   \cr\cr
   m_1 & m_2 & m_3}
   $$
   $$
   \times  
   (j_1 m_1   \vert j_1 a_1 \Gamma_1 \gamma_1 )^* \ 
   (j_2 m_2   \vert j_2 a_2 \Gamma_2 \gamma_2 )^* \
   (j_3 m_3   \vert j_3 a_3 \Gamma_3 \gamma_3 )^* 
   \eqno (4)
         $$
and
$$
  \pmatrix{
                        & j &                         \cr\cr
  a_1 \Gamma_1 \gamma_1 &   &a_2 \Gamma_2 \gamma_2\cr}
  \; := \; 
  \sqrt{2 j + 1} \; 
  \overline f 
  \pmatrix{
  j                     & 0                     & j\cr\cr
  a_1 \Gamma_1 \gamma_1 & a_0 \Gamma_0 \gamma_0 & a_2 \Gamma_2 \gamma_2}.
  \eqno (5)
$$
 On the right-hand side of (3), the quantum numbers $\tau_1$ 
and $\tau_2$, external to the chain $SU(2) \supset G^*$ (they 
depend on the physics of the considered problem), 
appear only in the reduced matrix element 
$(\tau_1 j_1 \Vert U^{(k)} \Vert \tau_2 j_2)$.  

The $\overline f$ or $3$-$j a \Gamma \gamma$ symbol in (4) is an 
$SU(2) \supset G^*$ symmetry adapted form of the usual $3$-$jm$ 
Wigner symbol. It constitutes a symmetrized form of the coefficient 
\cite{K68}
        $$
  f 
  \pmatrix{
  j_1                   & j_2                   & k\cr\cr
  a_1 \Gamma_1 \gamma_1 & a_2 \Gamma_2 \gamma_2 & a \Gamma \gamma}
  \;  :=  \;  \sum_{m_1 q m_2} \;
  (- 1)^{j_1 - m_1} \;
  \pmatrix{
  j_1  & k & j_2 
  \cr\cr
  -m_1 & q & m_2}
  $$
  $$ 
    \times  
    (j_1 m_1 \vert j_1 a_1 \Gamma_1 \gamma_1)^* \;
    (k   q   \vert k   a   \Gamma   \gamma)     \;
    (j_2 m_2 \vert j_2 a_2 \Gamma_2 \gamma_2)   
  \eqno (6)
        $$
that generalizes the $f$ coefficient defined via 
$$
  (J \Gamma \gamma \vert U^{(k)}_{\Gamma_0} \vert J' \Gamma' \gamma') 
  \; := \; 
  \delta (\Gamma', \Gamma)  \; 
  \delta (\gamma', \gamma)  \; 
  f
  \pmatrix{
  J      & J'     & k\cr\cr
  \Gamma & \Gamma & \Gamma_0}
  \eqno (7)
$$
in Mosh\'e's M.Sc. \cite{F60}.

The $2$-$j a \Gamma \gamma$ symbol defined by (5) 
turns out to be an $SU(2) \supset G^*$ symmetry 
adapted form of the Herring-Wigner 
metric tensor whose spherical 
components may be taken as 
$$
  \pmatrix{
      & j &     \cr \cr
  m_1 &   & m_2}
  \; := \; (-1)^{j + m_1} \; \delta (m_2, - m_1). 
  \eqno (8)
$$
From equations (5) and (8) we have
$$
  \pmatrix{
                        & j &                       \cr\cr
  a_1 \Gamma_1 \gamma_1 &   & a_2 \Gamma_2 \gamma_2}
  \; = \; \sum_{m} \ (-1)^{j + m} \
  (j  m \vert j a_1 \Gamma_1 \gamma_1)^* \;
  (j,-m \vert j a_2 \Gamma_2 \gamma_2)^*. 
  \eqno (9)
$$
The metric tensor given by (9) allows us to handle all 
the phases occurring in the $\left\{ j a \Gamma \gamma \right\}$ 
scheme. 

By combining (4), (6) and (9), we can rewrite (3) in the 
simple form
$$
  (\tau_1 j_1 a_1 \Gamma_1 \gamma_1 \vert U^{(k)}_{a \Gamma \gamma} \vert 
   \tau_2 j_2 a_2 \Gamma_2 \gamma_2) \; = \; 
  (\tau_1 j_1 \Vert U^{(k)} \Vert \tau_2 j_2)  \; 
  f \pmatrix{
  j_1                   & j_2                   & k\cr\cr
  a_1 \Gamma_1 \gamma_1 & a_2 \Gamma_2 \gamma_2 & a \Gamma \gamma}. 
  \eqno (10)
$$
The interest of (3) and (10) for electronic spectroscopy of ions in 
crystalline fields has 
been discussed at length in Ref.~\cite{K69IJQC}. From a mathematical 
viewpoint, it is to be observed that the factorization in 
(10) into the product of a $G^*$--dependent 
factor (the $f$ coefficient) 
by a $G^*$--independent factor (the reduced matrix element) is valid 
whether the group $G^*$ is multiplicity free or not. The internal 
multiplicity problem arising when $G^*$ is not 
multiplicity free is thus easily solved by making use of (10). 

An important property of the $f$ coefficient is 
given by the following factorization formula which 
arises as a consequence of a lemma by Racah \cite{RIV}. Indeed, we
have
        $$
  f \pmatrix{
  j_1 & j_2 & k\cr
  \cr
  a_1 \Gamma_1 \gamma_1&a_2 \Gamma_2 \gamma_2&a \Gamma \gamma} \; = 
  \; (-1)^{2k} \; \frac{1}{ \sqrt {2j_1 + 1} }                          
  $$
  $$
  \times \sum_{\beta} \; 
  (j_2 a_2 \Gamma_2 + k a \Gamma \vert j_1 a_1 \beta \Gamma_1)^* \ 
  (\Gamma_2 \Gamma \gamma_2 \gamma \vert 
   \Gamma_2 \Gamma \beta \Gamma_1 \gamma_1)^*, 
  \eqno (11)
        $$
where 
$(j_2 a_2 \Gamma_2 + k a \Gamma \vert j_1 a_1 \beta \Gamma_1)$ 
is an isoscalar factor or reduced  $f$  symbol 
(independent of the   labels   $\gamma_1$, $\gamma_2$ and $\gamma$) and 
$(\Gamma_2 \Gamma \gamma_2 \gamma \vert 
  \Gamma_2 \Gamma \beta \Gamma_1 \gamma_1)$ 
a Clebsch-Gordan coefficient for the group $G^*$ considered as 
an isolated entity. The label $\beta$ is an internal 
multiplicity label to be used when the Kronecker product 
$\Gamma_2 \otimes \Gamma$ is not multiplicity free. 

Equations (10) and (11)    are   of central importance in 
the study of  magnetic and electronic 
properties of an $\ell^N$ ion in $G$ 
(with $G \sim G^*/Z_2$) symmetry as well as in rotational 
spectroscopy of molecules invariant under $G$. They proved to be 
useful in the development of  phenomenological models  for  
dealing with 
both energy levels and intensities of transitions. The reader may 
consult Refs.~[23-26] for an application to crystal-field theory.

 \end{document}